\documentclass[12pt,preprint]{aastex}

\shortauthors{Acciari et al.}

\begin{document}


\title{Discovery of Very High-Energy Gamma-Ray Radiation\\from the BL Lac 1ES\,0806+524}

\author{
V. Acciari\altaffilmark{1},
E. Aliu\altaffilmark{2},
T. Arlen\altaffilmark{3},
M. Bautista\altaffilmark{4},
M. Beilicke\altaffilmark{5},
W. Benbow\altaffilmark{6},
M. B{\"o}ttcher\altaffilmark{7},
S. M. Bradbury\altaffilmark{8},
J. H. Buckley\altaffilmark{5},
V. Bugaev\altaffilmark{5},
Y. Butt\altaffilmark{9},
K. Byrum\altaffilmark{10},
A. Cannon\altaffilmark{11},
O. Celik\altaffilmark{3},
A. Cesarini\altaffilmark{12},
Y. C. Chow\altaffilmark{3},
L. Ciupik\altaffilmark{13},
P. Cogan\altaffilmark{4},
P. Colin\altaffilmark{14},
W. Cui\altaffilmark{15},
R. Dickherber\altaffilmark{5},
C. Duke\altaffilmark{16},
T. Ergin\altaffilmark{9},
A. Falcone\altaffilmark{17},
S. J. Fegan\altaffilmark{3},
J. P. Finley\altaffilmark{15},
G. Finnegan\altaffilmark{14},
P. Fortin\altaffilmark{18},
L. Fortson\altaffilmark{13},
A. Furniss\altaffilmark{19},
D. Gall\altaffilmark{15},
K. Gibbs\altaffilmark{6},
G. H. Gillanders\altaffilmark{12},
J. Grube\altaffilmark{11},
R. Guenette\altaffilmark{4},
G. Gyuk\altaffilmark{13},
D. Hanna\altaffilmark{4},
E. Hays\altaffilmark{20},
J. Holder\altaffilmark{2},
D. Horan\altaffilmark{21},
C. M. Hui\altaffilmark{14},
T. B. Humensky\altaffilmark{22},
A. Imran\altaffilmark{23},
P. Kaaret\altaffilmark{24},
N. Karlsson\altaffilmark{13},
M. Kertzman\altaffilmark{25},
D. Kieda\altaffilmark{14},
J. Kildea\altaffilmark{6},
A. Konopelko\altaffilmark{26},
H. Krawczynski\altaffilmark{5},
F. Krennrich\altaffilmark{23},
M. J. Lang\altaffilmark{12},
S. LeBohec\altaffilmark{14},
G. Maier\altaffilmark{4},
A. McCann\altaffilmark{4},
M. McCutcheon\altaffilmark{4},
J. Millis\altaffilmark{27},
P. Moriarty\altaffilmark{1},
R. Mukherjee\altaffilmark{18},
T. Nagai\altaffilmark{23},
R. A. Ong\altaffilmark{3},
A. N. Otte\altaffilmark{19},
D. Pandel\altaffilmark{24},
J. S. Perkins\altaffilmark{6},
D. Petry\altaffilmark{28},
M. Pohl\altaffilmark{23},
J. Quinn\altaffilmark{11},
K. Ragan\altaffilmark{4},
L. C. Reyes\altaffilmark{29},
P. T. Reynolds\altaffilmark{30},
E. Roache\altaffilmark{6},
J. Rose\altaffilmark{8},
M. Schroedter\altaffilmark{23},
G. H. Sembroski\altaffilmark{15},
A. W. Smith\altaffilmark{10},
D. Steele\altaffilmark{13},
S. P. Swordy\altaffilmark{22},
M. Theiling\altaffilmark{6},
J. A. Toner\altaffilmark{12},
L. Valcarcel\altaffilmark{4},
A. Varlotta\altaffilmark{15},
V. V. Vassiliev\altaffilmark{3},
R. G. Wagner\altaffilmark{10},
S. P. Wakely\altaffilmark{22},
J. E. Ward\altaffilmark{11},
T. C. Weekes\altaffilmark{6},
A. Weinstein\altaffilmark{3},
R. J. White\altaffilmark{8},
D. A. Williams\altaffilmark{19},
S. Wissel\altaffilmark{22},
M. Wood\altaffilmark{3},
B. Zitzer\altaffilmark{15}
}

\altaffiltext{1}{Department of Life and Physical Sciences, Galway-Mayo Institute of Technology, Dublin Road, Galway, Ireland}
\altaffiltext{2}{Department of Physics and Astronomy and the Bartol Research Institute, University of Delaware, Newark, DE 19716, USA}
\altaffiltext{3}{Department of Physics and Astronomy, University of California, Los Angeles, CA 90095, USA}
\altaffiltext{4}{Physics Department, McGill University, Montreal, QC H3A 2T8, Canada}
\altaffiltext{5}{Department of Physics, Washington University, St. Louis, MO 63130, USA}
\altaffiltext{6}{Fred Lawrence Whipple Observatory, Harvard-Smithsonian Center for Astrophysics, Amado, AZ 85645, USA}
\altaffiltext{7}{Astrophysical Institute, Department of Physics and Astronomy, Ohio University, Athens, OH 45701}
\altaffiltext{8}{School of Physics and Astronomy, University of Leeds, Leeds, LS2 9JT, UK}
\altaffiltext{9}{Harvard-Smithsonian Center for Astrophysics, 60 Garden Street, Cambridge, MA 02138, USA}
\altaffiltext{10}{Argonne National Laboratory, 9700 S. Cass Avenue, Argonne, IL 60439, USA}
\altaffiltext{11}{School of Physics, University College Dublin, Belfield, Dublin 4, Ireland}
\altaffiltext{12}{School of Physics, National University of Ireland, Galway, Ireland}
\altaffiltext{13}{Astronomy Department, Adler Planetarium and Astronomy Museum, Chicago, IL 60605, USA}
\altaffiltext{14}{Physics Department, University of Utah, Salt Lake City, UT 84112, USA}
\altaffiltext{15}{Department of Physics, Purdue University, West Lafayette, IN 47907, USA }
\altaffiltext{16}{Department of Physics, Grinnell College, Grinnell, IA 50112-1690, USA}
\altaffiltext{17}{Department of Astronomy and Astrophysics, 525 Davey Lab, Pennsylvania State University, University Park, PA 16802, USA}
\altaffiltext{18}{Department of Physics and Astronomy, Barnard College, Columbia University, NY 10027, USA}
\altaffiltext{19}{Santa Cruz Institute for Particle Physics and Department of Physics, University of California, Santa Cruz, CA 95064, USA}
\altaffiltext{20}{N.A.S.A./Goddard Space-Flight Center, Code 661, Greenbelt, MD 20771, USA}
\altaffiltext{21}{Laboratoire Leprince-Ringuet, Ecole Polytechnique, CNRS/IN2P3, F-91128 Palaiseau, France}
\altaffiltext{22}{Enrico Fermi Institute, University of Chicago, Chicago, IL 60637, USA}
\altaffiltext{23}{Department of Physics and Astronomy, Iowa State University, Ames, IA 50011, USA}
\altaffiltext{24}{Department of Physics and Astronomy, University of Iowa, Van Allen Hall, Iowa City, IA 52242, USA}
\altaffiltext{25}{Department of Physics and Astronomy, DePauw University, Greencastle, IN 46135-0037, USA}
\altaffiltext{26}{Department of Physics, Pittsburg State University, 1701 South Broadway, Pittsburg, KS 66762, USA}
\altaffiltext{27}{Department of Physics, Anderson University, 1100 East 5th Street, Anderson, IN 46012}
\altaffiltext{28}{European Southern Observatory, Karl-Schwarzchild-Strasse 2, 85748 Garching, Germany}
\altaffiltext{29}{Kavli Institute for Cosmological Physics, University of Chicago, Chicago, IL 60637, USA}
\altaffiltext{30}{Department of Applied Physics and Instumentation, Cork Institute of Technology, Bishopstown, Cork, Ireland}
\email{peter.cogan@mail.mcgill.ca}








\begin{abstract}
The high-frequency-peaked BL-Lacertae object \objectname{1ES 0806+524}, at redshift z=0.138, was observed in the very-high-energy
(VHE) gamma-ray regime by VERITAS between November 2006 and April
2008. These data encompass the two-, and three-telescope commissioning
phases, as well as observations with the full four-telescope
array. \objectname{1ES 0806+524} is detected with a statistical
significance of 6.3 standard deviations from 245 excess events. Little
or no measurable variability on monthly time scales is found. The
photon spectrum for the period November 2007 to April 2008 can be
characterized by a power law with photon index $3.6 \pm
1.0_{\mathrm{stat}} \pm 0.3_{\mathrm{sys}}$ between $\sim$300 GeV and
$\sim$700 GeV. The integral flux above 300 GeV is
$(2.2\pm0.5_{\mathrm{stat}}\pm0.4_{\mathrm{sys}})\times10^{-12}\:\mathrm{cm}^{-2}\:\mathrm{s}^{-1}$
which corresponds to 1.8\% of the Crab Nebula
flux. Non-contemporaneous multiwavelength observations are combined
with the VHE data to produce a broadband spectral energy distribution
that can be reasonably described using a synchrotron-self-Compton
model.
\end{abstract}





\section{Introduction}

Active galactic nuclei (AGN) are galactic cores with a luminosity that
dominates the host galaxy. These central engines are believed to
be powered by supermassive black holes which draw surrounding matter
into an accretion disc. The unified model \citep{1995PASP..107..803U}
of AGN implies that the observed emission from AGN is a strong
function of the observation angle in relation to the relativistic jet,
with the highest energy observed emission beamed along the direction
of the jet.

\objectname{1ES 0806+524} was identified as a BL Lacertae object
\citep{1993ApJ...412..541S} using radio observations from the Green
Bank 91-m telescope \citep{becker91} and X-ray observations from the
Einstein Slew Survey \citep{elvis92}. The redshift of the host galaxy
is z=0.138 \citep{bade98}. The blazar \objectname{1ES 0806+524} was
suggested as a candidate VHE gamma-ray source based on the
presence of both high-energy electrons and sufficient seed photons
\citep{costamante2002}. That work predicts an intrinsic flux in the
VHE regime of $\mathcal{F}_{\mathrm{E}>0.3\:\mathrm{TeV}}=1.36 \times
10^{-11}\:\mathrm{cm}^{-2}\:\mathrm{s}^{-1}$.

Very high-energy gamma-ray observations of \objectname{1ES 0806+524}
reported by the Whipple Collaboration
\citep{horan04a,delacalleperez03} indicated flux upper limits of
$\mathcal{F}_{\mathrm{E}>0.3\:\mathrm{TeV}}<1.4 \times
10^{-11}\:\mathrm{cm}^{-2}\:\mathrm{s}^{-1}$ and
$\mathcal{F}_{\mathrm{E}>0.3\:\mathrm{TeV}}<16.8 \times
10^{-11}\:\mathrm{cm}^{-2}\:\mathrm{s}^{-1}$ from the 1996 and 2000
observing seasons and $\mathcal{F}_{\mathrm{E}>0.3\:\mathrm{TeV}}<1.47
\times 10^{-11}\:\mathrm{cm}^{-2}\:\mathrm{s}^{-1}$ from the 2000 to
2002 observing seasons. Observations reported by the HEGRA
Collaboration \citep{2004A&A...421..529A} indicated an upper limit of
$\mathcal{F}_{\mathrm{E}>1.09\:\mathrm{TeV}}<43 \times
10^{-11}\:\mathrm{cm}^{-2}\:\mathrm{s}^{-1}$ based on hour of
observations. Observations of \objectname{1ES 0806+524} in late 2006
and early 2007 with VERITAS \citep{Cogan2007a} yielded evidence for
weak emission with a statistical significance of 2.5 standard
deviations in 35 hours. The data reported in \cite{Cogan2007a} are
combined with the data from the 2007/2008 observing season here.

\section{Observations and analysis}

VERITAS \citep{Weekes-2002} is an array of four 12-m diameter imaging
Cherenkov telescopes located at the Fred Lawrence Whipple Observatory
in southern Arizona (1268 m a.s.l.,~31$^{\mathrm{o}}40'30''$N,
$110^{\mathrm{o}}57'07''$W).

Observations of \objectname{1ES 0806+524} were made between November
2006 and February 2007 during the construction phase of VERITAS and
from November 2007 to April 2008 during the full four-telescope
operation. All observations were taken in \emph{wobble} mode
\citep{1994APh.....2..137F}, where the target is offset from the
center of the field of view by $\pm0.3^\circ$ or $\pm0.5^\circ$. After
quality selection to remove data suffering from poor weather or
technical problems, a total of 65 hours of observations are
available. These data comprise 5 hours with a T1/T2 configuration, 25
hours with a T1/T2/T3 configuration, and 35 hours with a T1/T2/T3/T4
configuration. These data were analyzed using the standard offline
VERITAS data analysis package {\sl VEGAS}
\citep{Cogan2007b}. Consistent results are obtained using independent
analysis packages.


Shower images of extensive air showers in each camera are
gain-corrected and cleaned \citep{Cogan2007c} before being
parameterised using a moment analysis
\citep{1985ICRC....3..445H}. Prior to stereoscopic reconstruction, a
pre-selection retaining images with size greater than 75
photoelectrons, with greater than 4 pixels surviving cleaning, and
with a distance from the center of the field of view to the image
centroid of less than $1.43^\circ$ is applied. Events where only
images from the two closest-spaced telescopes survive the
pre-selection are rejected. Each shower is parameterised using
mean-scaled width (MSW) and mean-scaled length (MSL)
\citep{Konopelko}. Gamma-ray selection criteria of $\mathrm{MSW} <
1.16$ and $\mathrm{MSL} < 1.36$ are used to reject most of the
background while retaining a large portion of the signal. These
selection criteria were optimised assuming a source strength of 3\% of
the Crab Nebula flux.

Background estimation is performed using the reflected-region
background model \citep{2007A&A...466.1219B}. In this scheme, an
integration region is placed around \objectname{1ES 0806+524}, with background
integration regions distributed around the field of view. The number
of events in the integration region is termed ON, the number of events
in the background region is termed OFF, and the ratio of the
integration areas is termed $\alpha$. As the data comprise
observations with different absolute wobble offsets, a generalized
version of equation 17 from \cite{1983ApJ...272..317L}, as noted in
\cite{2004A&A...421..529A}, is used to compute the statistical
significance of the excess.

The entire VERITAS dataset from 2006 through 2008 reveals the
existence of very-high-energy gamma-ray emission from \objectname{1ES
  0806+524} with a statistical significance of 6.3 standard deviations
above an energy threshold of $\sim$300 GeV. The total number of ON counts is
1543, the total number of OFF counts is 13311 and the effective ratio
of the integration regions (accounting for the different wobble
offsets) is $\alpha=0.0975$, yielding an excess of 245 events.

\begin{figure}
\epsscale{.80}
\plotone{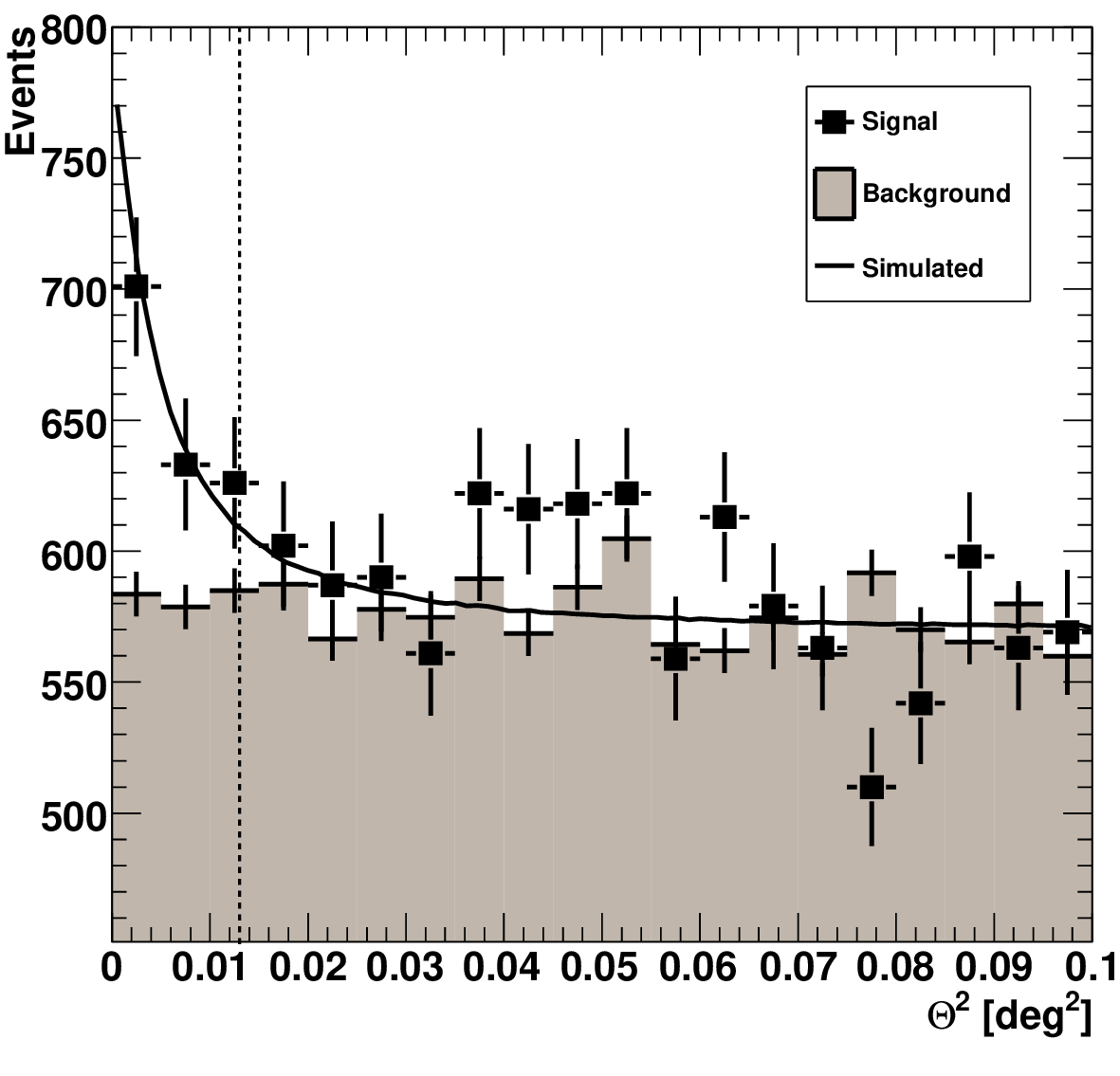}
\caption{ Distribution of the parameter
  $\theta^{2}$, the squared angular distance between the reconstructed
  shower direction and the location of 1ES\,0806+524. The vertical
  line, located at $\theta^{2}=0.013^{\circ^2}$ indicates the size of
  the integration region. The simulated line indicates the
  expected shape of the $\theta^{2}$ distribution for a point source
  of this strength.\label{fig:ThetaSquare}}
\end{figure}

A distribution of the squared angular distance (referred to as
$\theta^2$) between the position of \objectname{1ES 0806+524} and the
reconstructed shower direction is shown in Figure
\ref{fig:ThetaSquare}. The integration region used in the
reflected-region analysis is indicated by the vertical line situated
at $\theta^2=0.013^{\circ^2}$. The scaled background counts are shown
for comparison, and are computed using the centers of the reflected
regions as the reference points. The distribution of excess gamma-ray
counts is consistent with that given by a point source located at
right ascension $8^{\mathrm{h}} 9^{\mathrm{m}} 59^{\mathrm{s}} \pm
9^{\mathrm{s}}$ and declination $52^\circ 19' \pm 2'$. There is an
added combined systematic angular uncertainty of $\sim$100''. This is
consistent, within error, with the position of the radio source at
right ascension $8^{\mathrm{h}} 9^{\mathrm{m}} 49.2^{\mathrm{s}}$ and
declination $52^\circ 18' 58''$ \citep{2007AJ....133.1236K}.

Only those data taken with four operating telescopes are used for a
spectral analysis, owing to the superior energy resolution
afforded. The subset of data, taken during the 2007/2008 observing
period, reveals a total of 176 ON events, 1334 OFF events with
statistical significance of $4.4\:\sigma$ using equation 17 from
\cite{1983ApJ...272..317L}, where $\alpha=0.09$, yielding an excess of
55 events. The differential photon spectrum can be fit using a power law
of the form $\mathrm{dN/dE} = F_0\times\left(\mathrm{E/400
  GeV}\right)^{-\Gamma}$ where $F_0=(6.8 \pm 1.7_{\mathrm{stat}} \pm
1.3_{\mathrm{sys}}) \times 10^{-12} \mathrm{cm}^{-2} \mathrm{s}^{-1}$
and $\Gamma=3.6 \pm 1.0_{\mathrm{stat}} \pm 0.3_{\mathrm{sys}}$. The
fit yields a $\chi^2/\mathrm{dof}$ of 0.07/2, where dof is the number
of degrees of freedom. Upper limits for two spectral points above $\sim$700
GeV are calculated at the 90\% confidence level according to
\cite{1984NIMPA.228..120H}.

Absorption on the infrared component of the extragalactic background
light results in an attenuation of high-energy photons
\citep{1967PhRv..155.1408G}. The absorption model according to
\cite{2008A&A...487..837F} is used to calculate the deabsorbed
spectrum. This is achieved by scaling each flux point according to
$\mathcal{F}_\mathrm{int}=\mathcal{F}_\mathrm{obs}e^{\tau\left(E,z\right)}$,
where $E$ is the energy at that flux point, $z$ is the redshift and
$\tau$ is a function describing the absorption model. A power law fit
to the resultant flux points results in a spectral index of
$-2.8\pm0.5_{\mathrm{stat}}$.

\begin{figure}
\epsscale{.80}
\plotone{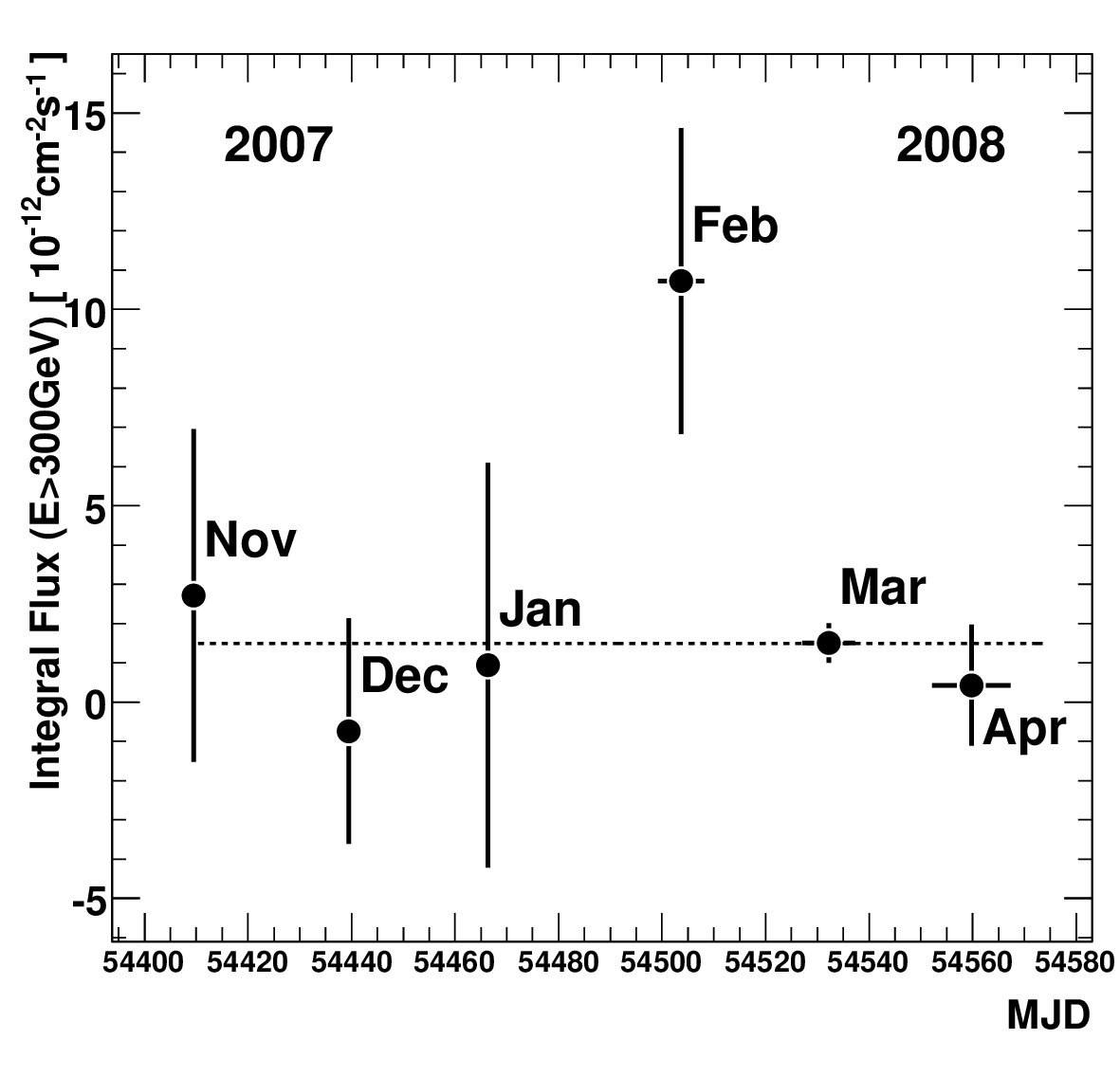}
\caption{Month-by-month integral flux above 300
  GeV from observations of 1ES\,0806+524. The chi-square probability
  of the straight line fit is 0.24.\label{fig:lightcurve}}
\end{figure}

The integral light curve above 300 GeV is shown in Figure
\ref{fig:lightcurve} with the data binned per month. A constant fit
yields a $\chi^2/\mathrm{dof}$ of 6.78/5 indicating little or no
measurable variability.









\section{Analysis of Swift UVOT and XRT Data}

Following the announcement of the discovery of 1ES 0806+524 by VERITAS
in the very-high energy gamma-ray regime \citep{2008ATel.1415....1S},
\objectname{1ES 0806+524} was observed by Swift \citep{2004ApJ...611.1005G} in
the ultraviolet to optical and X-ray energy bands using the UVOT and
XRT between 2008 March 8 and 2008 March 12. Data reduction is carried
out with the HEAsoft 6.4 package. The Swift XRT data were taken in
photon counting mode at a rate below 0.6 counts/s, so photon
pileup is not evident. The XRTPIPELINE tool is used to calibrate and
clean the XRT event files. Source and background counts are extracted
from circular regions of radius 30 and 40 pixels, respectively. The
0.6 to 10 keV energy spectra are fit by a power law with fixed
Galactic column density of N $_{\rm{H}} = 4.4 \cdot 10^{20}$ cm$^{-2}$
\citep{1990ARA&A..28..215D}. Marginal flux variability is seen from 2008 March 8 to 12,
with the photon index hardening from $\Gamma = 2.67 \pm 0.08$ to
$\Gamma = 2.53 \pm 0.07$.

Swift UVOT data is reduced with the UVOTSOURCE tool to extract
counts, correct for coincidence losses, apply background subtraction,
and calculate the source flux. The standard 5 arcsec radius source
aperture is used, with a 20 arcsec background region. The source
fluxes are dereddened using the interstellar extinction curve in
\cite{1999PASP..111...63F}.

\section{Discussion and Modeling}

\begin{figure}
\epsscale{.80}
\plotone{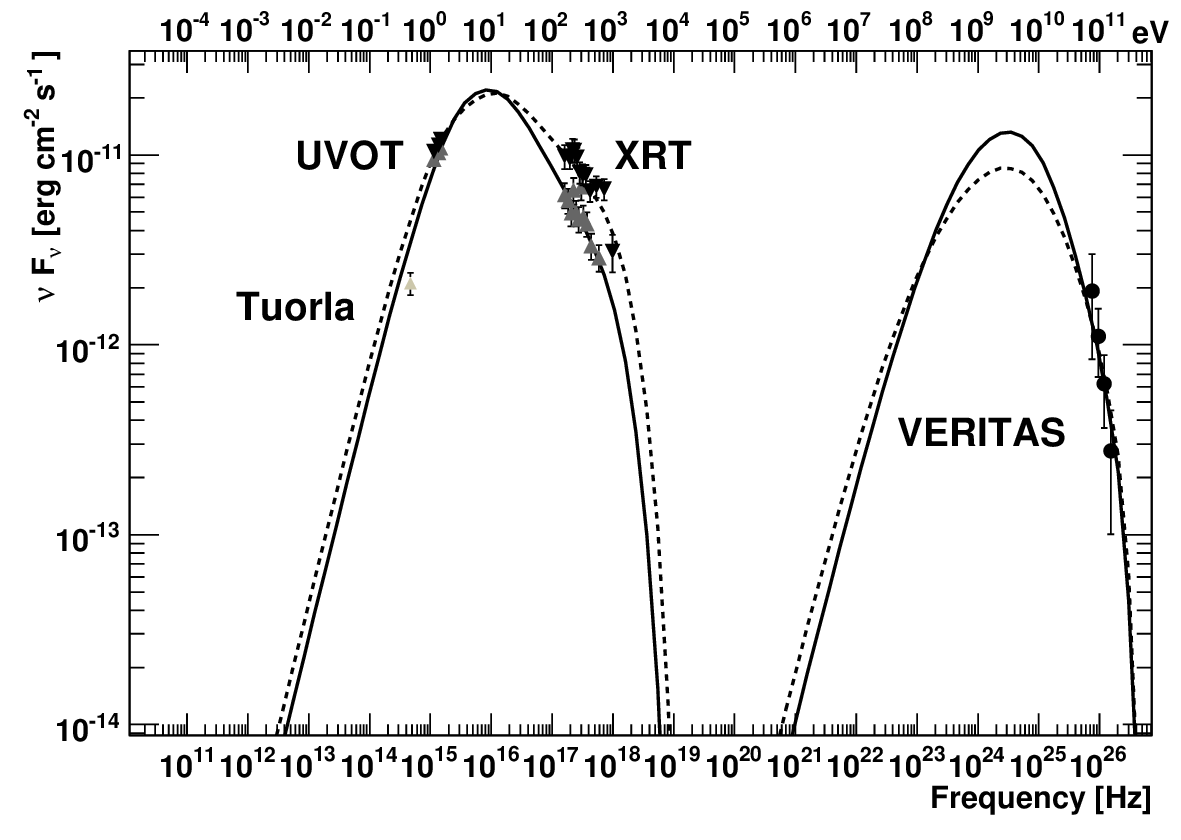}
\caption{Broadband spectral energy distribution of 1ES\,0806+524. The
  Swift data are taken on two separate nights, with a lower flux on
  2008 March 8 (gray points) and a higher flux on 2008 March 12 (black
  points). The solid and dashed curves are SSC fits to the 2008 March
  8 and 2008 March 12 Swift data, respectively.\label{fig:sed}}
\end{figure}

Non-contemporaneous data from the Swift UVOT\footnote[1]{Note that the
  host galaxy contribution is not important in UV.} and XRT are
combined with the VERITAS data and an archival optical data point
(from the Tuorla 1m
telescope\footnote[2]{http://users.utu.fi/kani/1m/}) to produce a
broadband spectral energy distribution in a $\nu F_\nu$ representation
as shown in Figure \ref{fig:sed}. The Tuorla data were taken with an
R-band filter and host-galaxy subtraction is applied. The spectral
energy distribution (SED) is modeled using the equilibrium version of
the one-zone jet radiation transfer code of
\cite{2002ApJ...581..127B}, with parameters appropriate for a pure
synchrotron-self-Compton (SSC) model. This model assumes a population
of ultrarelativistic nonthermal electrons (and positrons) injected
into a spherical comoving volume (referred to as the \emph{blob}). The
injection spectrum is characterized by an injection power
L$_\mathrm{inj}\left(t\right)$ with an unbroken power-law distribution
between energies $\gamma_1$ and $\gamma_2$ with index $q$. The jet
moves with relativistic speed $\beta$c and it is charaterized by
Doppler factor $D$ which is dependent on the line of sight. The
injected particles suffer radiative losses via synchrotron emission
and Compton upscattering of synchrotron photons.

The SSC model has been fitted to the data, with different parameters
obtained for the 2008 March 8 and 2008 March 12 data sets. For 2008
March 8, we obtain L$_\mathrm{inj}\left(t\right)
=1.9\times10^{43}\mathrm{\:erg\:s}^{-1}$ with an injection spectrum
described by $\gamma_1=1.77\times10^4$, $\gamma_2=2\times10^5$,
spectral index $q=3.1$ and a magnetic field strength of
$0.39\:\mathrm{Gauss}$. The corresponding parameters for the 2008
March 12 dataset are $1.6\times10^{43}\mathrm{\:erg\:s}^{-1}$,
$\gamma_1=1.6\times10^4$, $\gamma_2=2\times10^5$, injection spectral
index $q=2.7$ and magnetic field strength of
$0.5\:\mathrm{Gauss}$. The blob radius is
$5\times10^{15}\:\mathrm{cm}$ in both cases. A Lorentz factor of
$\Gamma=20$ is used, with the simplifying assumption that the Doppler
factor D is equal to the Lorentz factor. The synchrotron peak is
located at $\sim$8.3$\times10^{15}$ Hz and the inverse-Compton peak is
located at $\sim$3.5$\times10^{24}$ Hz. The system is within a factor
$\sim$2 of equipartition in both cases. Absorption on the
extragalactic background light using the EBL scenario described by
\cite{2008A&A...487..837F} has been accounted for in the model.





\section{Conclusions}

VERITAS has observed the blazar \objectname{1ES 0806+524} for a total
65 hours from November 2006 to April 2008 resulting in the discovery
of very-high-energy gamma rays with a statistical significance of
$6.3\:\sigma$. The differential energy spectrum between $\sim$300 GeV
and $\sim$700 GeV can be fitted by a relatively soft power law with
index $\Gamma=3.6 \pm 1.0_{\mathrm{stat}}\pm 0.3_{\mathrm{sys}}$. The
integral flux above 300 GeV is
$(2.2\pm0.5_{\mathrm{stat}}\pm0.4_{\mathrm{sys}})\times10^{-12}\:\mathrm{cm}^{-2}\:\mathrm{s}^{-1}$
which corresponds to $\left(1.8\pm0.5\right)$\% of the Crab Nebula
flux as measured by VERITAS above 300 GeV. Assuming absorption on the
infrared component of the extragalactic background light according to
\cite{2008A&A...487..837F}, the intrinsic integral flux above 300 GeV
is
$(4.4\pm0.6_{\mathrm{stat}}\pm0.5_{\mathrm{sys}})\times10^{-12}\:\mathrm{cm}^{-2}\:\mathrm{s}^{-1}$
which is approximately one
order of magnitude less than the flux predicted by
\cite{costamante2002}. The broadband spectral energy distribution can
be fitted using a one-zone SSC model with standard parameters.

Observations by the EGRET gamma-ray space telescope reveal gamma-ray
emission from the region around \objectname{1ES 0806+524}, however the
large error box indicates that the emission could be associated with
either \objectname{B 0803+5126} (z=1.14) or \objectname{1ES 0806+524},
with the former being favored
\citep{2003ApJ...590..109S}. Observations of this region with the
Large Area Telescope on the Fermi Gamma-Ray Space Telescope should
resolve the source(s) of gamma-ray emission in this region.

VERITAS is the most sensitive instrument of its kind in the Northern
Hemisphere, and it is ideally suited to observations of extragalactic
objects. This has been demonstrated by the detection of three new BL
Lac objects by VERITAS in 2008 (\cite{2008ATel.1415....1S},
\cite{2008ATel.1753....1S}, \cite{2008ApJ...684L..73A}).



\acknowledgments

This research is supported by grants from the U.S. Department of
Energy, the U.S. National Science Foundation, the Smithsonian Institution,
by NSERC in Canada, by Science Foundation Ireland and by PPARC in the
UK. We acknowledge the excellent work of the technical support staff
at the FLWO and the collaborating institutions in the construction and
operation of VERITAS.

The authors thank the anonymous referee for helpful comments which
helped to improve and clarify the text.



{\it Facilities:} \facility{VERITAS}, \facility{Swift}




\clearpage



\begin{figure}
\epsscale{.80}
\plotone{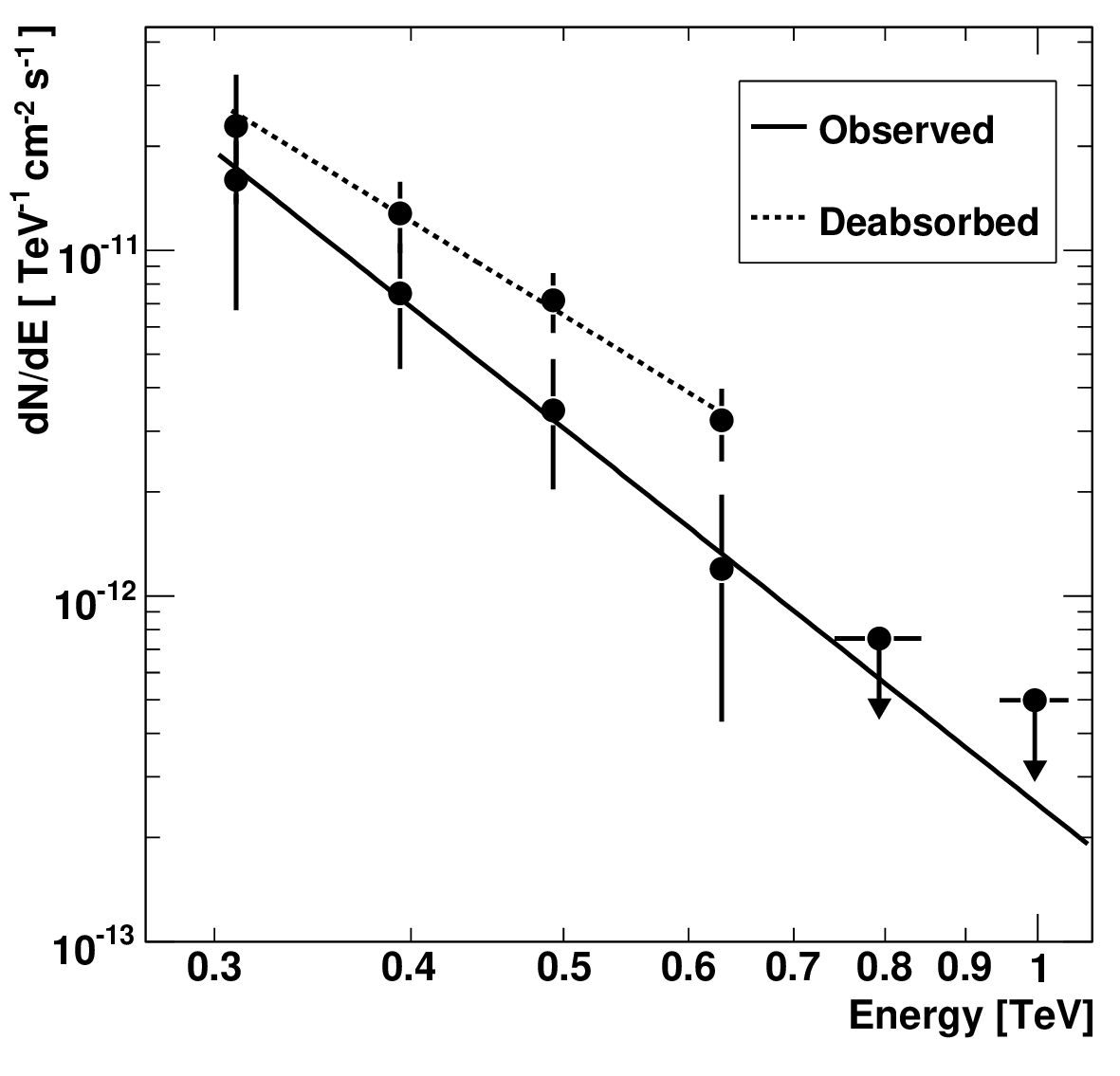}
\caption{Differential photon spectrum of 1ES\,0806+524. The spectrum
  is well fit by a power law with index $ 3.6 \pm 1.0_{\mathrm{stat}}
  \pm 0.3_{\mathrm{stat}}$. The deabsorbed spectrum is calculated by
  applying the extragalactic absorption model according to
  \cite{2008A&A...487..837F}. \label{fig:Spectrum}}
\end{figure}

\end{document}